\documentclass[epj,nopacs]{svjour}
\newif\ifpdf\ifx\pdfoutput\undefined\pdffalse\else\pdfoutput=1\pdftrue\fi
\newcommand{\pdfgraphics}{\ifpdf\DeclareGraphicsExtensions{.pdf, .jpg}\else\DeclareGraphicsExtensions{.eps, .ps, .jpg}\fi}

\usepackage{graphicx,amssymb,euscript}
\begin{document}
\pdfgraphics

\title{Factorization and Sudakov Resummation in {\boldmath $B\to\gamma l\nu$\unboldmath}}
\author{Stefan W. Bosch}
\institute{Newman Laboratory for Elementary-Particle Physics, Cornell University, Ithaca, NY 14853, U.S.A. \hfill CLNS 03/1842}
\date{Received: date / Revised version: date}
%
\abstract{
We apply Soft-Collinear Effective Theory to prove at leading power in $\Lambda_{\mathrm{QCD}}/m_{b}$ a factorization formula for the radiative leptonic decay $B\to\gamma l\nu$. Large logarithms entering the hard-scattering kernel are systematically resummed by a two-step perturbative matching procedure. 
\PACS{
      {PACS-key}{discribing text of that key}   \and
      {PACS-key}{discribing text of that key}
     } 
} 
\maketitle
\section{Introduction}
\label{sec:intro}
One of the main goals in today's $B$ physics is to understand exclusive $B$ decays in order to perform accurate CKM studies and New Physics searches. Major steps in this direction were taken recently using heavy-quark expansions. The QCD factorization formalism \cite{BBNS} allows in the heavy-quark limit $m_{b}\gg\Lambda_{\mathrm{QCD}}$ a separation of perturbatively calculable hard-scattering kernels from nonperturbative form factors and universal light-cone distribution amplitudes. The ``Soft-Collinear Effective Theory'' (SCET) \cite{SCET} for the strong interactions of collinear and soft particles allows us to better understand the factorization properties of hard exclusive processes and eventually rigorously prove the corresponding factorization theorems. In particular for hadronic and radiative $B$-meson decays into light particles a challenge is to understand the interactions of collinear particles with the soft spectator quark inside the $B$ meson which gives rise to convolutions of hard-scattering kernels with the $B$-meson light-cone distribution amplitude (LCDA). A systematic analysis of these interactions in the framework of SCET has recently been performed in \cite{HN}.

The decay $B\to \gamma l\nu$ provides a particularly clean environment for the study of soft-collinear interactions because no hadrons are present in the final state yet the light-cone structure of the $B$ meson is probed by the coupling of the high-energy photon to the soft spectator quark inside the $B$ meson. Furthermore, three mass scales are involved in the process which lead to Sudakov logarithms. We use the formalism of \cite{HN} to prove a factorization formula for $B\to\gamma l\nu$ and systematically resum the large Sudakov logarithms. Related work on $B\to\gamma l\nu$ was presented at this conference by Descotes-Genon and Lunghi \cite{Bgamlnu}. We present our results of \cite{BHLN}. Other aspects are discussed in \cite{Bgamlnuother}.

\section{Soft-Collinear Effective Theory}
\label{sec:SCET}
SCET is an effective field theory of collinear particles interacting with soft degrees of freedom. It borrows ideas from HQET, NRQCD, and collinear effective theory and is based on the method of regions, i.e. leading regions correspond to effective fields. Fluctuations with $p^{2} \gtrsim Q^{2}\gg\Lambda_{\mathrm{QCD}}^{2}$ are integrated out and appear in Wilson coefficients whereas those with $p^{2}\ll Q^{2}$ appear in time-ordered products of effective theory fields. The main advantage of an effective field theory approach is that symmetries of the theory are explicit at the level of the Lagrangian, which simplifies factorization proofs. 

In the formulation of SCET by Hill and Neubert \cite{HN}, soft and collinear fields appear as gauge-invariant building blocks living on light-like trajectories. These building blocks absorb the SCET Wilson lines such that gauge invariance is no longer a constraint on the form of the SCET operators. Yet, reparameterization invariance \cite{RPI} gives powerful constraints.

\section{The factorization formula and its proof}
\label{sec:fact}
For a high-energy photon with $E_{\gamma}={\cal O}(m_{b})$ we have the factorization formula \cite{Bgamlnu,BHLN}
\begin{eqnarray}\label{ff}
   \lefteqn{{\cal A}(B^-\to\gamma\,l^-\bar\nu_l) }\\
   && \propto m_B f_B\,Q_u \int_0^\infty\!dl_+\, \frac{\phi_+^B(l_+,\mu)}{l_+}\,T(l_+,E_\gamma,m_b,\mu)\nonumber
\end{eqnarray}
where $Q_u=\frac23$ is the electric charge of the up-quark , $f_B$ is the $B$-meson decay constant, $\phi_+^B$ is a leading-order LCDA of the $B$-meson, and $T=1+O(\alpha_s)$ is a perturbative hard-scattering kernel. The physics underlying the factorization formula is that a high-energy photon coupling to the soft constituents of the $B$-meson produces quantum fluctuations far off their mass shell, which can be integrated out in a low-energy effective theory. 

Let us briefly sketch the four ingredients of the proof of (\ref{ff}) to all orders in perturbation theory. We match the full theory amplitude onto the unique set of gauge-invariant SCET operators that mediate this decay and that are allowed by reparameterization invariance. Then we show that (see \cite{BHLN} for further details)
\begin{itemize}
\item The decay amplitude can be, at leading power, expressed in terms of a convoution with the $B$ meson LCDA because the component field of the effective operators have light-like separation.
\item The hard scattering kernel is free of infrared divergences to all orders in perturbation theory because it is a SCET Wilson coefficient.
\item The convolution integral of the hard-scattering kernel with the $B$-meson LCDA is convergent  using reparameterization invariance arguments.
\item Non-valence Fock states give no leading-power contributions because additional soft gluon fields are either power suppressed or vanishing.
\end{itemize}

\section{Hard-scattering kernel and resummation}
\label{sec:kernel}
Because the hard-scattering kernel is a SCET Wilson coefficient we calculate it by a matching procedure. To next-to leading order in QCD we find
\begin{eqnarray}\label{Tres}
   \lefteqn{T(l_+,E_\gamma,m_b,\mu) = 1 + \frac{C_F\,\alpha_s(\mu)}{4\pi} \bigg[  - 2\ln^2\frac{2E_\gamma}{\mu} - \frac{\pi^2}{4} } \\
&& + \ln^2\frac{2E_\gamma\,l_+}{\mu^2} \nonumber - \frac{x_\gamma\ln x_\gamma}{1-x_\gamma} + 2\ln\frac{2E_\gamma}{\mu} - 2L_2(1-x_\gamma) -5 \bigg] \nonumber
\end{eqnarray}
$T$ contains large logarithms which cannot be made small simultaneously for any choice of $\mu$. However, we can resum these large logarithms using a two-step matching onto SCET. In a first step the ${\cal O}(m_{b}^{2})$ off-shell fluctuations of the $b$ quark are integrated out by matching onto HQET. This yields a hard function $H$. Hard-collinear modes with momenta scaling like $(k_+,k_-,k_\perp) \sim(\Lambda,m_b,\sqrt{m_b\Lambda})$ are integrated out in a second step and lead to the jet function $J$. We thus find a second stage of perturbative factorization of the hard scattering kernel
\begin{equation}\label{fact2}
   T(l_+,E_\gamma,m_b,\mu)  = H\left(\frac{2E_\gamma}{\mu},\frac{2E_\gamma}{m_b}\right) \cdot J\left(\frac{2E_\gamma\,l_+}{\mu^2}\right)
\end{equation}
where, at NLO,
\begin{eqnarray*}
   H\left(\frac{2E_\gamma}{\mu},x_\gamma\right)
   &=& 1 + \frac{C_F\,\alpha_s(\mu)}{4\pi}
   \bigg[ - 2\ln^2\frac{2E_\gamma}{\mu} + 2\ln\frac{2E_\gamma}{\mu}\\
   &&- \frac{x_\gamma\ln x_\gamma}{1-x_\gamma} - 2L_2(1-x_\gamma)
   - 4 - \frac{\pi^2}{12} \bigg]
   \phantom{\Bigg|}\\
   J\left(\frac{2E_\gamma\,l_+}{\mu^2}\right)
   &=&  1 + \frac{C_F\,\alpha_s(\mu)}{4\pi}
    \bigg( \ln^2\frac{2E_\gamma\,l_+}{\mu^2} - 1 - \frac{\pi^2}{6}
    \bigg)
\end{eqnarray*}
We actually performed the matching diagrammatically and determined $H$ simply by setting the soft spectator momentum in the $B$ meson $l=0$. The jet function then followed from the ratio $J=T/H$.

To get the renormalization group equation for $T$ we use the scale independence of the total amplitude and the renormalization properties of the LCDA $\phi_{+}^{B}(\omega ,\mu)$ to get the integro-differential equation \cite{LN} 
\begin{eqnarray}
   \frac{d}{d\ln\mu}\,T(l_+,\mu) &=& \left[ \Gamma_{\rm cusp}(\alpha_s)\,\ln\frac{\mu}{l_+} + \gamma(\alpha_s) \right] T(l_+,\mu) \nonumber\\
   && + \int_0^\infty\!d\omega\,l_+\,\Gamma(\omega,l_+,\alpha_s)\, T(\omega,\mu) \nonumber
\end{eqnarray}
where $\Gamma_{\rm cusp}$ is the universal cusp anomalous dimension 
familiar from the theory of the renormalization of Wilson loops \cite{Gammacusp}. Its appearance is due to the fact that a $B$ meson can be described by a Wilson line $S_{n}(z,0)S_{v}(0,-\infty)$ with a cusp singularity at the origin which gives rise to one factor of $\Gamma_{\rm cusp}$ \cite{LN}.

From the factorization property of the hard-scattering kernel exhibited in (\ref{fact2}) and the functional forms of the hard and jet functions given above, it follows that the hard component and the jet function obey the RG equations
\begin{eqnarray*}
   \frac{d}{d\ln\mu}\,H(\mu)
   &=& \left[ - \Gamma_{\rm cusp}\,\ln\frac{\mu}{2E_\gamma}
    + \gamma(\alpha_s) - \gamma'(\alpha_s) \right] H(\mu)  \\
   \frac{d}{d\ln\mu}\,J(l_+,\mu)
   &=& \left[ \Gamma_{\rm cusp}\,
    \ln\frac{\mu^2}{2E_\gamma\,l_+} + \gamma'(\alpha_s) \right]
    J(l_+,\mu)\\
    && + \int_0^\infty\!d\omega\,l_+\,
    \Gamma(\omega,l_+,\alpha_s)\,J(\omega,\mu)
\end{eqnarray*}
where explicit expressions for $\Gamma$, $\gamma$, and $\gamma'$ can be found in \cite{BHLN}. The reasoning in deriving the RG equations for $H$ and $J$ is analogous to an argument presented by Korchemsky and Sterman in their discussion of the $B\to X_s\gamma$ photon spectrum \cite{Korchemsky:1994jb}. 
We note that only a single logarithm of $\mu/2 E_{\gamma}$ appears in the RG equations so that it is possible to integrate them. Furthermore the coefficient of the logarithm is the cusp anomalous dimension which is known to two-loop order. This allows for the resummation of Sudakov logarithms at NLO.

To perform the Sudakov resummation we first calculate the hard function $H(\mu_{h})$ at a high scale $\mu_{h}\sim m_{b}$ using fixed-order perturbation theory. At this scale no large logarithms are present. Solving the RGE for $H$ we can evolve $H(\mu)$ down to an intermediate scale $\mu_{i}\sim\sqrt{m_{b}\Lambda}$ and multiply it by the result $J(l_{+},\mu_{i})$ for the jet function. Again, no large logarithms appear and we get the kernel $T(l_{+},\mu_{i})$ at the intermediate scale. Finally, we solve the RGE for $T$ and compute the evolution down to a low-energy scale $\mu\sim\mbox{few}\times\Lambda_{\mathrm{QCD}}$. The exact solution for the resummed kernel is given by
\[
   T(l_+,\mu) = H(\mu_h)\,{\EuScript J}[\alpha_s(\mu_i),\nabla_\eta]\,
   \exp U(l_+,\mu,\mu_i,\mu_h,\eta) \Big|_{\eta=0}
\]
where we refer to \cite{BHLN} for the definition of ${\EuScript J}[\alpha_s(\mu_i),\nabla_\eta]$ and the explicit expression for the evolution function $U$.

\begin{figure*}
\center{\resizebox{1.0\textwidth}{!}{%
  \includegraphics{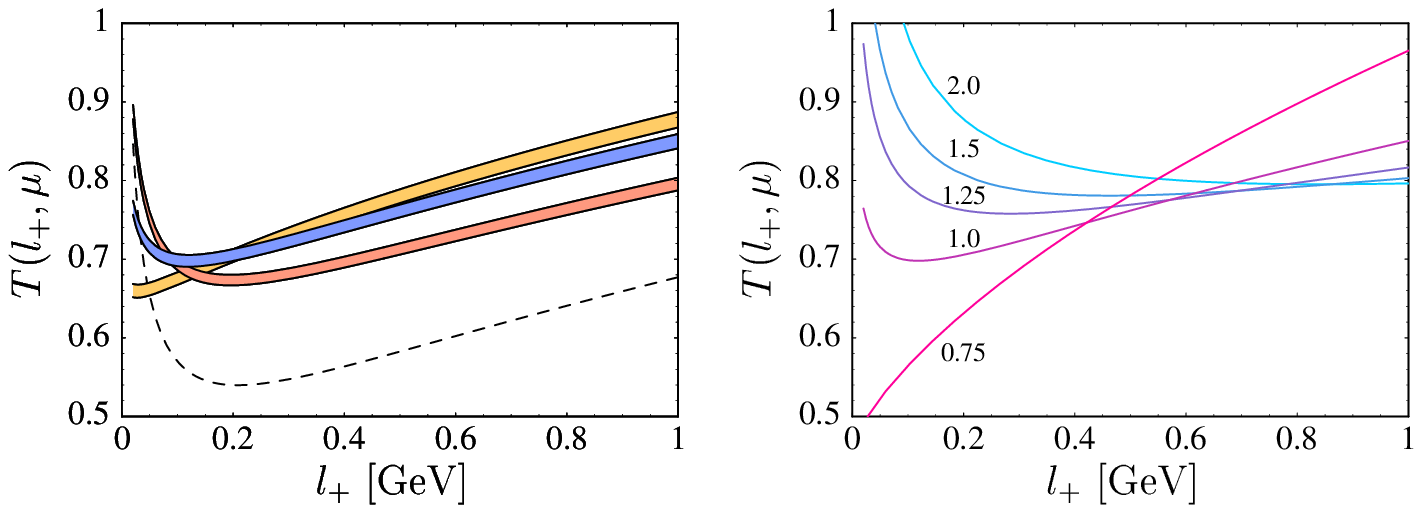}\includegraphics{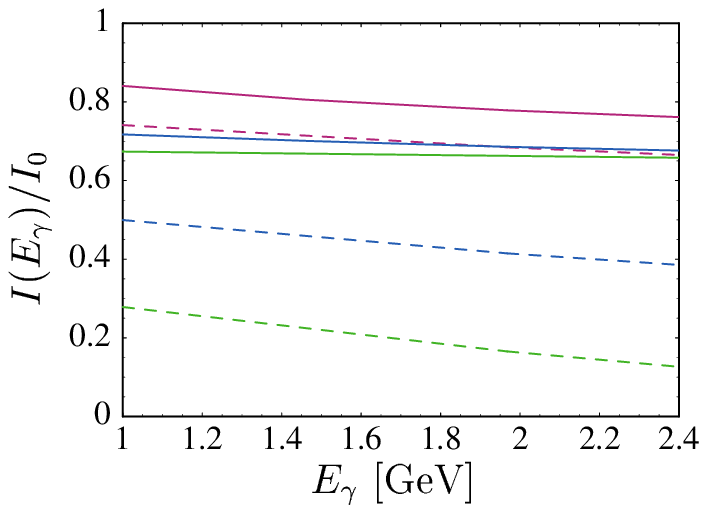}
}}
  \caption{\label{fig:resumTI} RG-improved predictions for the hard-scattering kernel at maximum photon energy and the convolution integral $I(E_{\gamma})$.}
\end{figure*}
In order to study the importance of RG improvement and Sudakov resummation, we compare in the left-hand plot in Figure~\ref{fig:resumTI} the result for the resummed hard-scat\-ter\-ing kernel at maximal photon energy $E_{\gamma}=m_{b}/2$ with the one-loop approximation in (\ref{Tres}). We plot the function $T(l_{+},\mu=1\mathrm{GeV})$ for different choices of the matching scales $\mu_{i}$ (different bands) and $\mu_{h}$ (width of bands). The dashed line shows the result obtained at one-loop order without resummation. We find that resummation effects decrease the magnitude of the radiative corrections, i.e., the resummed kernel is closer to the tree-level value $T=1$ than the one-loop result. Our results are stable under the variation of $\mu_{i}$ and $\mu_{h}$.

The scale dependence of the resummed expression for the kernel is illustrated in the middle plot in Figure~\ref{fig:resumTI}, which shows the functional dependence of $T(l_+,\mu)$ for maximal photon energy and several values of $\mu$. The matching scales are set to their default values $\mu_h=m_b=4.8$\,GeV and $\mu_i=\sqrt{\Lambda_h m_b}\simeq 1.55$\,GeV. We observe a significant scale dependence of the kernel, especially as one lowers $\mu$ below the intermediate scale $\mu_i$. In other words, the second stage of running (for $\mu<\mu_i$) is numerically significant.

Finally, the right-hand plot in Figure~\ref{fig:resumTI} shows the resummation effects for the convolution integral
\[
   I(E_\gamma) = \int_0^\infty\!dl_+\,\frac{\phi_+^B(l_+,\mu)}{l_+}\, T(l_+,E_\gamma,m_b,\mu)
\]
in units of its tree-level value $I_{0}=1/\lambda_{B}$ for three different values of the hadronic scale $\mu_{0}$ at which we assumed the following particular form of the LCDA
\[
\phi_+^B(l_+,\mu_0) = \frac{l_+}{\lambda_B^2}\,e^{-l_+/\lambda_B}
\]
with $\lambda_B = \frac23\,(m_B-m_b)\approx 0.32\,\mbox{GeV}$. This ansatz was motivated by a QCD sum-rule analysis \cite{GN}. With respect to its tree-level value we observe a modest reduction of $I(E_\gamma)$ after RG resummation (solid lines) which is fairly insensitive to the precise value of $\mu_0$ and only shows a mild energy dependence. In contrast, the results obtained at one-loop order (dashed curves) are strongly sensitive to the choice of $\mu_0$ and exhibit a more pronounced dependence on the photon energy.

\section{Conclusions}\label{sec:concl}
We have applied soft-collinear effective theory to prove a QCD factorization formula for the radiative semileptonic decay $B\to\gamma l\nu$ to all orders in perturbation theory. We showed that, at leading power in $\Lambda_{\mathrm{QCD}}/m_{b}$, the amplitude can be written as a convolution of a perturbative, infrared-finite hard-scattering kernel with the leading-order $B$-mes\-on LCDA. Additionally, we have shown that the convolution integral is free of endpoint singularities and that non-valence Fock states of the $B$ mes\-on do not contribute at leading power.

Furthermore we presented the calculation of the hard-scattering kernel in the factorization formula using re\-nor\-mal\-ization-group improved perturbation theory. We have established a second perturbative factorization formula according to which the different short-distance scales entering in the calculation of the kernel can be separated into a hard function and a jet function. The corresponding two classes of large logarithms can be systematically resummed by solving evolution equations derived from the renormalization properties of the leading-order $B$-meson light-cone distribution amplitude. We found that Sudakov resummation does not lead to a strong suppression of the decay amplitude.

The discussion of the decay $B\to\gamma l\nu$ presented here can be taken over almost verbatim to analyze related processes such as $B\to\gamma\gamma$ and $B\to\gamma\,l^+ l^-$. 

{\em Acknowledgement:} This research was supported by the National Science Foundation under Grant PHY-0098631.

\end{document}